\documentclass[conference]{IEEEtran}
\usepackage{amsmath,amssymb,graphicx,epstopdf,enumitem,pgfplots,cite}
\usepackage{algorithm,algorithmic}
\usepackage[nocomma]{optidef}
\usepackage[caption=false,font=footnotesize]{subfig}
\usepackage{balance}
\usetikzlibrary{shapes,arrows,positioning,patterns}

\long\def\comment#1{}

\newfont{\bbb}{msbm10 scaled 700}

\newfont{\bb}{msbm10 scaled 1100}

\newcommand{\rv}{{\bf r}}

\newcommand{\Am}{{\bf A}}

\makeatletter
\def\footnoterule{\relax%
  \kern-5pt
  \hbox to \columnwidth{\vrule width 0.5\columnwidth height 0.4pt\hfill}
  \kern4.6pt}
\makeatother

\newcommand\blfootnote[1]{%
  \begingroup
  \renewcommand\thefootnote{}\footnote{#1}%
  \addtocounter{footnote}{-1}%
  \endgroup
}

\newenvironment{varalgorithm}[1]
  {\algorithm[tb!]\renewcommand{\thealgorithm}{#1}}
  {\endalgorithm}

\newlength\figureheight
\newlength\figurewidth

\begin{document}

\title{A Distributed Approach for Networked Flying Platform Association with Small Cells\\ in 5G+ Networks}

\author{\IEEEauthorblockN{Syed Awais W. Shah\IEEEauthorrefmark{1},
Tamer Khattab\IEEEauthorrefmark{1},
Muhammad Zeeshan Shakir\IEEEauthorrefmark{2},
Mazen O. Hasna\IEEEauthorrefmark{1}}
\IEEEauthorblockA{\IEEEauthorrefmark{1}Department of Electrical Engineering, Qatar University, Doha, Qatar\\ Emails: \{syed.shah, tkhattab, hasna\}@qu.edu.qa}
\IEEEauthorblockA{\IEEEauthorrefmark{2}School of Engineering and Computing, University of the West of Scotland, Paisley, Scotland, UK\\
Email: muhammad.shakir@uws.ac.uk}}


\maketitle

\begin{abstract}
The densification of small-cell base stations in a 5G architecture is a promising approach to enhance the coverage area and facilitate the ever increasing capacity demand of end users. However, the bottleneck is an intelligent management of a backhaul/fronthaul network for these small-cell base stations. This involves efficient association and placement of the backhaul hubs that connects these small-cells with the core network. Terrestrial hubs suffer from an inefficient non line of sight link limitations and unavailability of a proper infrastructure in an urban area. Seeing the popularity of flying platforms, we employ here an idea of using networked flying platform (NFP) such as unmanned aerial vehicles (UAVs), drones, unmanned balloons flying at different altitudes, as aerial backhaul hubs. The association problem of these NFP-hubs and small-cell base stations is formulated considering backhaul link and NFP related limitations such as maximum number of supported links and bandwidth. Then, this paper presents an efficient and distributed solution of the designed problem, which performs a greedy search in order to maximize the sum rate of the overall network. A favorable performance is observed via a numerical comparison of our proposed method with optimal exhaustive search algorithm in terms of sum rate and run-time speed.\\
\end{abstract}


\section{Introduction}\label{sec:Intro}
\blfootnote{This publication was made possible by the sponsorship agreement in support of research and collaboration by Ooredoo, Doha, Qatar. The statements made herein are solely the responsibility of the authors.}
Fifth-generation (5G) will be a paradigm shift, where high data rate and wider coverage will be handled by the introduction of network facility in all components of a communication system. One example of such a network facility is heterogenous networks (HetNets), which includes densification of small-cell base stations (SBSs) (e.g., pico and femto cells), to cater the explosive growth of users \cite{Paradigm_Andrew, 5G_Andrew}. The idea is to bring the users closer to the BSs in order to make their association more reliable and thus, to satisfy the deluge of data rate demand of the overall network. In such a network, consisting of a large number of SBSs, a major challenge is their connectivity with the core network, known as backhaul/fronthaul \cite{robson2012backhaul}.

Backhaul hub routes the traffic between SBSs and core network using either wired (e.g., fiber optics) \cite{Fiber_Backhaul} or wireless (e.g., microwave or millimeter wave (mmWave)) technologies \cite{FSO_Ter_backhaul}. Wired optical fiber connection is always the best option but not the best choice due to its high capital expenditure (CAPEX). Wireless backhaul links can be categorized into line of sight (LoS) and non line of sight (NLoS) cases. LoS backhaul links use free space optics (FSO) or mmWave/microwave bands that leads to less coverage due to short range communication. On the other hand, NLoS radio frequency (RF) backhaul links do not have the coverage issue but they suffer from low data rate and hub placement problems because of few available ground locations in an urban area. Recently, a cost effective and a scalable idea of replacing the terrestrial backhaul network with an aerial network is presented in \cite{ShakirFSOMAG}, which employs networked flying platform (NFP) such as unmanned aerial vehicles (UAVs), drones, unmanned balloons, as aerial backhaul hubs. These NFP-hubs are meant for wireless communication as they are capable of communicating signalling and control information. Further, they hover at an altitude ranging from few hundred meters up to 20 kms including low altitude platforms (LAPs), medium altitude platforms (MAPs) and high altitude platforms (HAPs), depending upon coverage area, weather conditions and other related factors. Here, we use the idea of employing NFPs as backhaul hubs and present an efficient distributed algorithm for the association of multiple aerial NFP-hubs with the ground SBSs.

\begin{figure*}[tb!]\centering
    \includegraphics[width=11.5cm]{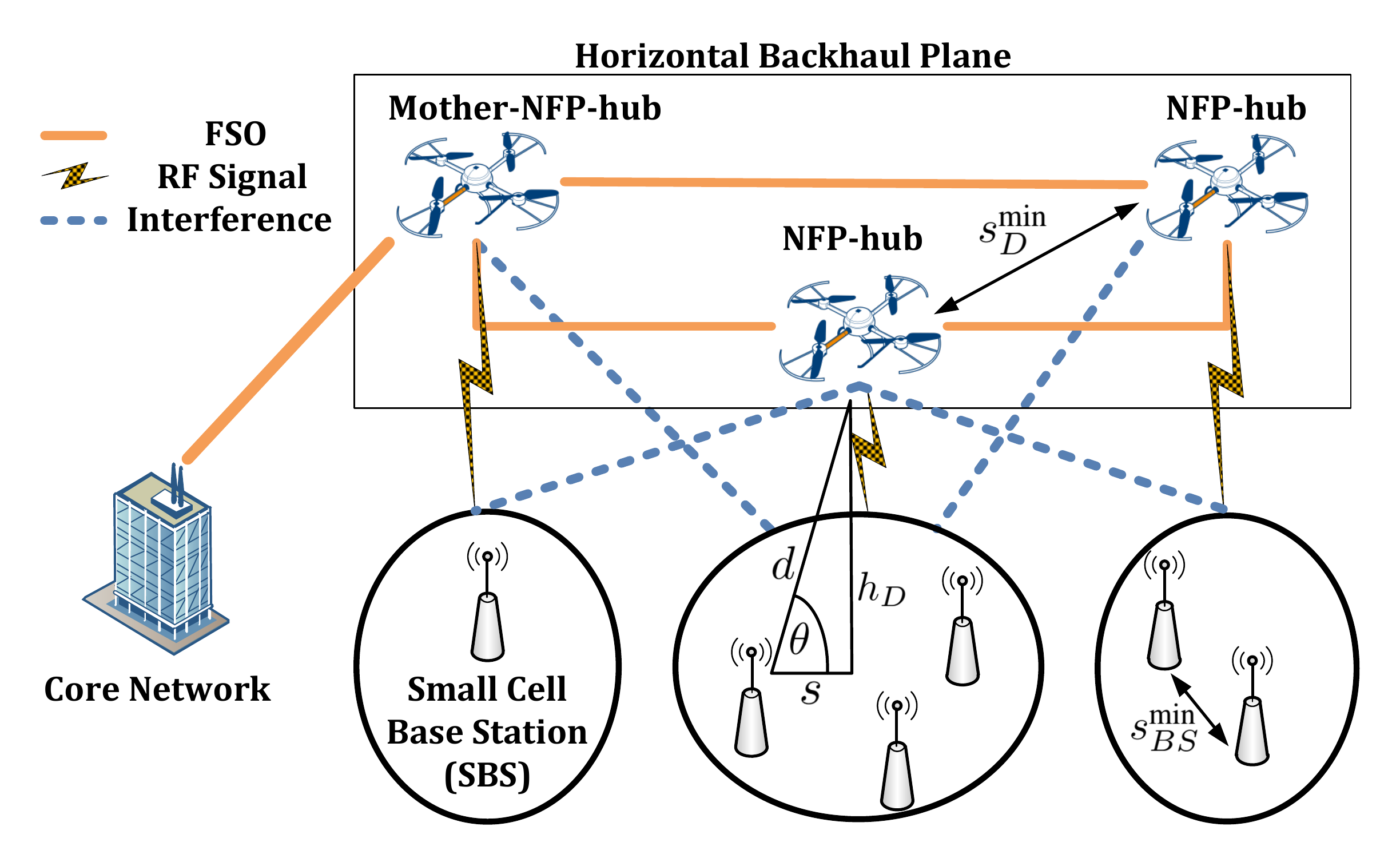}
	\caption{Graphical illustration of network flying platform and small-cell base station association problem.}
	\label{fig:SysMod}
\end{figure*}

\subsection{Related Work}\label{sec:RelWork}
Within a short span of time, the idea of using NFPs as relays and SBSs have attracted an eye of various researchers. The air to ground propagation model was presented in \cite{ATGmodel} for communication between LAP such as UAVs and terrestrial nodes. Then later on, a closed form expression was presented for this model and considering a fixed path loss (PL), an optimal altitude of a single UAV is analytically derived to maximize radio coverage \cite{ATG_optDrone1}. In \cite{TwoDrones}, the geographical coverage area is optimized for the case of two UAVs, considering their heights and distance between them as optimization parameters.

The issues of placement and association of UAVs with ground nodes were targeted by a few researchers \cite{IremOneDrone, ElhamBackhaul, ElhamMultiPSO, Mozaffari2016, Sharma2016}. In \cite{IremOneDrone}, the efficient placement of a single UAV as a BS was studied for different urban environments, satisfying minimum signal to noise ratio (SNR) as a qaulity of service (QoS) measure. A more comprehensive placement and association problem of a single UAV-BS was presented in \cite{ElhamBackhaul}, where a number of constraints including maximum PL, backhaul data rate and UAV bandwidth limit were considered. However, both \cite{IremOneDrone, ElhamBackhaul} solve the respective optimization problems using exhaustive search, which is not practically applicable. The case of multiple UAVs was considered in \cite{ElhamMultiPSO} where the number of UAV-BSs was computed considering serving and required capacities of UAV-BSs and ground users, respectively. Then, UAV-BSs and users were associated on the basis of the best signal-to-interference-plus-noise ratio (SINR), finally these UAV-BSs were placed by solving the optimization problem using particle swarm optimization (PSO) algorithm. This meta heuristic PSO algorithm requires a good initialization and a number of iterations to converge. In \cite{Mozaffari2016} and \cite{Sharma2016}, multiple UAV-BSs were deployed by solving the optimization problem considering only the SINR criterion and urban environment by utilizing optimal packing theory and game theory, respectively.

\subsection{Our Contributions}\label{sec:OurCont}
Most of the work done related to NFPs, employs them as either flying relays or BSs to enhance the network coverage and other parameters. Currently, there is only one work \cite{ShakirFSOMAG}, which investigates the feasibility of using NFPs as backhaul hubs but it was limited to designing backhaul framework, studying the effect of weather conditions and evaluation of the implementation cost of the proposed system. To the best of our knowledge, this is the first article which designs the association problem of NFP-hubs and SBSs and provides an efficient distributed greedy solution of the optimization problem contrary to the exhaustive search used in a number of related articles. Further, we have incorporated a number of practical constraints in our optimization problem, that were not used before, such as considering interference between NFP-hubs and SBSs using SINR parameter, maximum number of links that the NFP-hub can support because of the limited number of carried transceivers, maximum bandwidth supported by each NFP-hub and maximum backhaul data rate. Moreover, as opposed to related literature of NFPs, we have used a practical stochastic geometry approach for the random distribution of SBSs by keeping a minimum distance between them. Our designed algorithm is named as Distributed Maximal Demand Minimum Servers ((DM)$^2$S) as it is distributed among SBSs, NFP-hubs and mother-NFP-hub. This distribution enhances the run-time speed and performs a greedy search giving priority to SBSs demanding high data rate. Further, it maintains minimum links between NFP-hubs and SBSs to not overload NFP-hubs. Numerical results show a favorable performance of our proposed algorithm. Moreover, it is practically applicable and cost effective as compared to the exhaustive search.

The rest of the paper is organized as follows: In Section \ref{sec:SysMod}, a system model of NFP-hubs and SBSs considering backhaul framework is presented and the optimization problem for their association is designed. The proposed algorithm to solve the designed optimization problem is presented in Section \ref{sec:OptAlgo}. Section \ref{sec:SimRes} presents numerical analysis and discussions on the performance of the proposed method and exhaustive search. The computational complexity of algorithms is discussed in Section \ref{sec:Comp_Complexity} and Section \ref{sec:Conc} concludes the paper.

\section{System Model and Problem Formulation}\label{sec:SysMod}
Consider a HetNet as shown in Fig. \ref{fig:SysMod} consisting of three wireless nodes: i) ground SBSs, ii) NFP-hubs, and iii) ground core network. SBSs aggregate and route the uplink/downlink traffic of cellular users via backhaul NFP-hubs to the core network. NFP-hubs are connected to ground SBSs through a wireless RF link and these NFP-hubs are spread over a horizontal backhaul plane at a height $h_D$ from ground level. They are connected to each other through free space optical links\footnote{We assume a perfect LoS between NFP-hubs, thus no losses as well as no data rate/bandwidth limits are considered for these links (however, they may get affected by weather conditions \cite{ShakirFSOMAG}, that can be considered in future).} \cite{ShakirFSOMAG} and communicate with the core network through a mother-NFP-hub, which directly connects to the core network through another FSO link. Furthermore, we assume that the NFP-hubs are allowed to exchange control information with each other, however, every NFP-hub should directly transfer its data to the mother-NFP-hub. The control information includes the SINR of every NFP-hub to SBS links and bandwidth and data rate requirements of the SBSs. Moreover, we only consider the active SBSs during time interval $\begin{bmatrix} 0 & T \end{bmatrix}$ and during this time interval the system does not change. Below, we present air-to-ground (ATG) PL model and then formulate the problem.

\subsection{Air-to-Ground Path Loss Model}\label{sec:ATG_Model}
We adopt here a widely used ATG PL model presented in \cite{ATGmodel} and \cite{ATG_optDrone1}. This model considers two propagation groups: i) LoS receivers, and ii) NLoS receivers, where NLoS signals includes reflections and diffractions only. The probability of LoS is an important factor, which is based on the environment and the orientation of NFP-hubs and ground SBSs and it is formulated in \cite{ATGmodel} and \cite{ATG_optDrone1} as
\begin{equation}\label{Prob_LoS}
    P(\text{LoS}) = \frac{1}{1 + \alpha \exp \left\{ -\beta \left( \frac{180}{\pi} \theta - \alpha \right) \right\}}
\end{equation}
where $\alpha$ and $\beta$ are constants whose values depend on the environment (rural, urban, or others) and $\theta = \arctan \left( \frac{h_D}{s} \right)$ is the elevation angle from the ground SBS to the NFP-hub, where $s = \sqrt{ \left( x - x_D \right)^2 + \left( y - y_D \right)^2 }$ is the horizontal distance between them. The locations of SBSs and the NFPs in a Cartesian coordinate system is given as $\left(x,y\right)$ and $\left(x_D,y_D,h_D\right)$, respectively. The average PL is presented as
\begin{equation}\label{PathLoss}
  \begin{aligned}
    \text{PL}(dB) = &10 \log \left( \frac{4 \pi f_c d}{c} \right)^\gamma + P(\text{LoS}) \eta_{\text{LoS}}\\
                    &+ P(\text{NLoS}) \eta_{\text{NLoS}}
  \end{aligned}
\end{equation}
where the first term represents free space path loss (FSPL), which depends on carrier frequency $f_c$, speed of light $c$, PL exponent $\gamma$ and the distance $d=\sqrt{h_D^2+s^2}$ between NFP-hub and SBS. Variables $\eta_{\text{LoS}}$ and $\eta_{\text{NLoS}}$ represent additional losses for LoS and NLoS links, respectively and $P(\text{NLoS}) = 1 - P(\text{LoS})$; all of them depend on the environment. It can be noticed from \eqref{Prob_LoS} that the probability of LoS increases with the increase in the elevation angle and for a fixed PL and known distribution in \eqref{PathLoss}, one can estimate a geographical area covered by a NFP-hub relative to its height \cite{ATG_optDrone1}.

\subsection{Problem Formulation}\label{sec:Prob_Form}
Consider a downlink transmission from the core network to $N_{BS}$ SBSs through $N_D$ NFP-hubs. Considering a stochastic geometry approach, both SBSs and NFP-hubs are distributed randomly using \emph{Matern} type-I hard-core process \cite{matern2013} with an average density of $\lambda$ per $\text{m}^2$ having a minimum separation of $s_{BS}^{\text{min}}$ and $s_D^{\text{min}}$ with their neighbors, respectively. This provides a random distribution points of SBSs and NFP-hubs denoted as $\left(x_i,\,y_i\right)$ and $(x_{D_j}, y_{D_j}, h_{D_j})$, respectively, where $i \in \left\{ 1, \ldots, N_{BS}  \right\}$ and $j \in \left\{ 1, \ldots, N_D  \right\}$.

This communication is limited by a number of factors including maximum backhaul data rate $R$ of the link between the core network and mother-NFP-hub, maximum bandwidth $B_j$ of each NFP-hub available for SBSs, and maximum number of links $N_{l_j}$ that every NFP-hub can support. Furthermore, NFP-hub to SBS link should satisfy the QoS requirement depending upon the minimum SINR criterion. Here, we consider a snapshot of the SBSs and accordingly assume that the data rate and other requirements remain same for a small duration $T$, for which the position of the NFP-hubs remains fixed.

Our objective is to find the best possible association of the SBSs with the NFP-hubs such that the sum-rate of the overall system is maximized depending on a number of factors including $R$, $B_j$, $N_{l_j}$ and minimum SINR. Such a problem can be formulated as
\begin{subequations}
\label{eq:Opt_Prob}
\begin{alignat}{3}
    \max_{\{A_{ij}\}} \quad \sum_{i=1}^{N_{BS}} &\sum_{j=1}^{N_D} r_{ij} \cdot A_{ij} \label{Obj_Fun}\\
    \intertext{subject to}
    \sum_{i=1}^{N_{BS}} \sum_{j=1}^{N_D} r_{ij} \cdot A_{ij} \; &\leq \; R \label{cons1}\\
    \sum_{i=1}^{N_{BS}} b_{ij} \cdot A_{ij} \; &\leq \; B_j, && \quad \forall j \label{cons2}\\
    \text{SINR}_{ij} \cdot A_{ij} \; &\geq \; \text{SINR}_{\text{min}}, && \quad \forall i,j \label{cons3}\\
    \sum_{i=1}^{N_{BS}} A_{ij} \; &\leq \; N_{l_j}, && \quad \forall j \label{cons4}\\
    \sum_{j=1}^{N_D} A_{ij} \; &\leq \; 1, && \quad \forall i \label{cons5}
\end{alignat}
\end{subequations}
where optimization parameter $A_{ij}$ denotes the association of SBS with the NFP-hub as
\begin{equation}\label{Assoc_Mat}
      A_{ij} =
        \left\{
          \begin{array}{ll}
            1, & \hbox{if SBS $i$ connected with NFP-hub $j$,} \\
            0, & \hbox{otherwise.}
          \end{array}
        \right.
\end{equation}

The wireless backhaul link from the core network to the mother-NFP-hub limits the maximum allowed data rate of the entire network, which includes the total communication traffic from the SBSs or alternatively the NFP-hubs. This constraint is formulated as \eqref{cons1}, where $r_{ij}$ is the requested data rate of SBS $i$ associated with NFP-hub $j$. Moreover, in our system model, the data rate demand is distributed among SBSs randomly from a pre-defined data rate vector $\rv_{\text{SBS}}$. Thus, each SBS demands the same data rate from the NFP-hubs, i.e., $r_{ij} = r_i, \; \forall j$.

Constraint \eqref{cons2} represents the limit of the maximum bandwidth $B_j$ that the NFP-hub $j$ can distribute. This limit is associated with the wireless link of the second hop, i.e., from each NFP-hub to the connected SBSs. Here, $b_{ij} = \frac{r_{ij}}{\eta_{ij}}$  is the bandwidth available to each SBS $i$ connected with a NFP-hub $j$ and it is dependent on the demanded data rate $r_{ij}$ and spectral efficiency $\eta_{ij} = \log_2 \left( 1+\text{SINR}_{ij} \right)$, where SINR can be expressed as
\begin{equation}\label{eq:SINR}
      \text{SINR}_{ik} = \frac{P_{r_{ik}}}{\sum_{j=1,j \neq k}^{N_D} P_{r_{ij}} + \sigma}
\end{equation}
where $P_{r_{ij}}$ denotes the received power from NFP-hub $j$ to the SBS $i$ and $\sigma$ represents the noise floor of the link.

Constraint \eqref{cons3} ensures that every link from NFP-hub to SBS satisfies the required QoS requirement depending on the minimum SINR of the system. It plays a major role in association of SBSs with NFP-hubs, as minimum SINR results in maximum PL. Further, it can be noticed from \eqref{Prob_LoS}, that for a fixed PL, positions of SBSs and height of NFP-hubs $h_D$, we get a certain coverage area to be served by a NFP-hub \cite{ATG_optDrone1,TwoDrones}. Thus, each NFP-hub can serve the SBSs present in this particular coverage area. NFP-hub $j$ is capable of maintaining a maximum of $N_{l_j}$ links, which is included in constraint \eqref{cons4}. Further, constraint \eqref{cons5} restricts each SBS to be associated with only one NFP-hub.

\section{Optimization Algorithms}\label{sec:OptAlgo}
For a fixed location of NFP-hubs, the optimization problem \eqref{eq:Opt_Prob} is a Binary Integer Linear Program (BILP), which involves only the association problem of NFP-hubs with SBSs. Even for this association problem, satisfying the constraints \eqref{cons1} to \eqref{cons5} is very complicated in general and it is well known that there exists no standard method to solve such a NP-hard problem \cite{NPhard1, NPhard2}. Thus, below we use Branch and Bound (B\&B) algorithm, which is an exhaustive search, as an optimal benchmark solution. Next, we present our proposed algorithm, which is a simple but efficient greedy approach to solve the considered problem.

\subsection{Optimal Solution}\label{sec:OptSol}
The B\&B algorithm \cite{BandB_Algo} sets out all the possible solutions in the form of a rooted tree. Then, it examines the tree branches and estimates an upper and lower bounds of the optimal solution. There are a number of tools which uses B\&B algorithm such as CPLEX solver, MOSEK solver and MATLAB built-in integer linear program solver. Here, we utilize the MATLAB built-in solver to use B\&B method for an optimal solution. We use this optimal solution as a benchmark in comparison with our proposed method. However, such a B\&B method is computationally complex and expensive. Here, we compare the computational complexity in terms of elapsed time of the algorithms.

\subsection{Proposed Distributed Greedy Algorithm}\label{sec:HeuAlgo}
In the initialization step, we compute the required number of NFP-hubs and their distribution in a specified region. Here, we assume symmetry for all the NFP-hubs i.e., $h_{D_j}=h_D$, $B_j=B$ and $N_{l_j}=N_l$, however, our algorithm is applicable for the general case of optimization problem \eqref{eq:Opt_Prob} with necessary modifications. Moreover, it is assumed that the following information is available as a system parameter which includes the maximum number of links $N_l$ and bandwidth $B$ that each NFP-hub can support, the total number of SBSs $N_{BS}$ and their demanded data rate $r_{ij}$. In order to provide connectivity to every SBS, we first compute the number of SBSs that can be connected with a single NFP-hub $N_{BS}^D$, which is either defined as $N_l$ or as
\begin{equation}\label{NumDrones1}
  N_{BS}^D = \lfloor \frac{B}{b_{\text{avg}}} \rfloor
\end{equation}
where $b_{\text{avg}} = \frac{\sum_{i=1}^{N_{BS}} r_i}{N_{BS} \eta_{\text{avg}}}$ is the average bandwidth required by a SBS, $\eta_{\text{avg}}$ is the average spectral efficiency of the system and $\lfloor \cdot \rfloor$ denotes the floor function. Now, the total number of required NFP-hubs is computed as
\begin{equation}\label{NumDrones2}
  N_D = \lceil \frac{N_{BS}}{\min\{N_l,N_{BS}^{D}\}} \rceil
\end{equation}
where $\lceil \cdot \rceil$ represents the ceil function.

\begin{varalgorithm}{Initialization}
\small
\caption{System Initialization}
\label{algo:Intial}
\begin{algorithmic}[1]
    \REQUIRE $\lambda, \; \text{Area}, \; s_{BS}^{\text{min}}, \; h_{\text{max}}, \; PL_{\text{max}}, \; \alpha, \; \beta, \; \eta_{\text{LoS}}, \; \eta_{\text{NLoS}}$
    \ENSURE $\left(x_i,\, y_i\right), \; (x_{D_j}, y_{D_j}, h_{D_j})$
    \STATE \textbf{Distribution of SBSs:}
    \STATE $(x_i, y_i) \leftarrow$ \emph{Matern} Process$(Area, \lambda, s_{BS}^{\text{min}})$
    \STATE $N_{BS} \leftarrow$ Number of points in $(x_i, y_i)$
    \STATE \textbf{Distribution of NFP-hubs:}
    \STATE Compute $s_D^{\text{min}}$ using \eqref{Prob_LoS}, \eqref{PathLoss}, $PL_{\text{max}}, \alpha, \beta, \eta_{\text{LoS}}, \eta_{\text{NLoS}}$
    \STATE Compute $N_D$ using \eqref{NumDrones2}
    \STATE $(t_{x_i}, t_{y_i}) \leftarrow$ \emph{Matern} Process$(Area, \lambda, s_D^{\text{min}})$
    \STATE $(x_{D_j}, y_{D_j}) \leftarrow N_D$ points out of $(t_{x_i}, t_{y_i})$ and $h_{D_j} = h_{\text{max}}$
\end{algorithmic}
\end{varalgorithm}

The next step is to place these $N_D$ NFP-hubs so that they can cover a pre-defined area where $N_{BS}$ SBSs are placed. For this, we fix the height $h_D$ of every NFP-hub to a maximum allowed height denoted as $h_{D_{\text{max}}}$. Now, we compute the distance $s_D^{\text{min}}$ covered by a single NFP-hub for a fixed PL using \eqref{PathLoss}. Then, we distribute the NFP-hubs using \emph{Matern} type-I hard-core process with a minimum separation between them equal to $s_D^{\text{min}}$. Thus, at this point, we have the 3D locations $(x_{D_j}, y_{D_j}, h_D)$ of the NFP-hubs distributed randomly in a specified region. These steps are summarized in Algorithm \ref{algo:Intial}.

Next, we present below a greedy method to efficiently solve the association problem. This method is divided into following three steps.

\subsubsection{\textbf{Step 1}}\label{sec:Step1Sol}
The NFP-hubs send a broadcast initialization signal and each SBS computes the SINR using \eqref{eq:SINR} for its link with every NFP-hub. Every SBS selects the maximum SINR out of $N_D$ SINR values and also validates if this selected value is greater than minimum SINR as per constraint \eqref{cons3}. Then, it sends feedback with 1 to the selected NFP-hub corresponding to the maximum SINR and 0 feedback to others. Mathematically, we can say that up to this point, our association matrix $\Am$ has a number of entries with 1 corresponding to maximum SINR values for every NFP-hub to SBS link. Thus, for every SBS $i$ in a row of $\Am$, we have only a single non-zero entry for the selected NFP-hub $j$, which satisfies the constraint \eqref{cons5}. Note that, we used this maximum SINR methodology just to simplify the problem. At the end, each NFP-hub has a number of association requests from SBSs. Mathematically, for every $j^{\text{th}}$ NFP-hub in a column, there are a number of non-zero entries in association matrix $\Am$ corresponding to maximum SINR links.

\begin{varalgorithm}{(DM)$^2$S}
\small
\caption{Distributed Maximal Demand Minimum Servers Algorithm}
\label{my_algo}
\begin{algorithmic}[1]
    \REQUIRE $N_{BS}, \; N_D, \; N_l, \; B, \; R, \; \text{SINR}_{ij}, \; r_{ij}, \; b_{ij}$
    \ENSURE $\Am$
    \STATE Initialize: $\Am = \emptyset$
    \STATE \textbf{Step 1:} \hfill ($N_{BS} (N_D - 1)$)
    \FOR {$i = 1$ \TO $N_{BS}$}
        \STATE Select NFP-hub $j$ with max. $\text{SINR}_{ij}$
    \ENDFOR
    \STATE \textbf{Step 2:} \hfill ($N_l (N_{BS} - 1) + 4 N_l + 2$)
    \FOR {$j = 1$ \TO $N_D$}
        \STATE Initialize counters: $T_{N_l} = 0$, $T_b = 0$
        \WHILE {$T_{N_l} < N_l$ $\wedge$ $T_b < B$}
            \STATE Find max $r_{ij}$ with min. $b_{ij}$
            \IF {$T_b + b_{ij} \leq B$}
                \STATE Update $A_{ij}=1$, $T_{N_l} = T_{N_l}-1$ and $T_b = T_b + b_{ij}$
            \ENDIF
        \ENDWHILE
    \ENDFOR
    \STATE \textbf{Step 3:} \hfill ($N_{BS} (N_D - 1) + N_D + N_l +2$)
    \STATE Initialize: $T_r$ as total data rate of associated SBSs
    \WHILE {$T_r > R$}
        \STATE Select NFP-hub having min. associated links
        \STATE Select SBS with min. data rate
        \STATE De-associate selected NFP-hub to SBS pair as $A_{ij} = 0$
        \STATE Update total data rate as $T_r = T_r - r_{ij}$
    \ENDWHILE
\end{algorithmic}
\end{varalgorithm}

\subsubsection{\textbf{Step 2}}\label{sec:Step2Sol}
In this step, every NFP-hub selects a number of SBSs such that it tries to maximize the sum-rate and also to satisfy maximum bandwidth and links constraints \eqref{cons2} and \eqref{cons4}, respectively. As each NFP-hub, at this step, performs action on its own received list of SBSs's requests, thus, it can be performed distributively in order to save the elapsed time.

Every NFP-hub $j$ goes through its list and selects the SBS that requested for maximum data rate. Then, it updates the number of links and sum bandwidth counters and matches them with the maximum allowed links limit $N_l$ and bandwidth limit $B$, respectively. If the constraints are satisfied, it keeps that SBS $i$, otherwise discard its request by modifying $A_{ij}=0$ and move to the next SBS. Also, note that for SBSs requesting a same data rate, NFP-hub gives priority to the SBS requiring minimum bandwidth as per their link.

At the end of this step, for every NFP-hub $j$, we have a maximum of $N_l$ SBSs associated. Until now, we have fulfilled the objective criterion of maximizing the sum-rate and also taken care of the constraints \eqref{cons2} to \eqref{cons5} except the backhaul data rate constraint \eqref{cons1}, which we deal with in the next step.

\subsubsection{\textbf{Step 3}}\label{sec:Step3Sol}
Now, all the NFP-hubs share their association list with the mother-NFP-hub, which ensures the maximum backhaul data rate constraint \eqref{cons1} in the following manner. If the total rate of the associated SBSs $T_r$ satisfies $T_r < R$, then the algorithm completes. Otherwise, mother-NFP-hub selects some SBSs for de-association. For this, it searches for the NFPs associated with the minimum number of SBSs and starts de-associating their links first. For the selected NFP, it searches for the SBS $i$ requiring minimum data rate and then de-associates it if $T_r - r_{ij} \geq R$, otherwise it selects the SBS with the next higher data rate. Note that, in the entire algorithm, we give priority to SBSs demanding high data rate which is known as user centric case as defined in \cite{ElhamBackhaul}. After each de-association, mother-NFP-hub compares the total sum-rate with the backhaul data rate limit. If all of the links of the NFP-hub are dropped, it is considered as unused and thus, we update the number of drones as $N_D = N_D - 1$. Now, if still the backhaul data rate is not satisfied then mother-NFP-hub moves to the next NFP and repeats the procedure.

\begin{table}[tb!]
\renewcommand{\arraystretch}{1.3}
\centering
\caption{Simulation Parameters}
\label{tab:SimPar}
\begin{tabular}{|c|c|c|c|}
\hline
\textbf{Parameter}          & \textbf{Value} & \textbf{Parameter}       & \textbf{Value}   \\ \hline
$\alpha$                    & 9.61           & $\beta$                  & 0.16             \\ \hline
$\eta_{\text{LoS}}$         & 1 dB           & $\eta_{\text{NLoS}}$     & 20 dB            \\ \hline
$f_c$                       & 2 GHz          & $P_t$                    & 5 Watts          \\ \hline
$\text{SINR}_{\text{min}}$  & -5 dB          & $\text{PL}_{\text{max}}$ & 110 dB           \\ \hline
$R$                         & 2 Gbps         & $B$                      & 250 MHz          \\ \hline
$N_l$                       & 7              & $h_{D_{\text{max}}}$     & 300 meters       \\ \hline
$\rv_{\text{SBS}}$            & \multicolumn{3}{c|}{ \{ 30, 60, 90, 120, 150 \} Mbps }       \\ \hline
\end{tabular}
\end{table}

This algorithm provides an efficient solution of the optimization problem \eqref{eq:Opt_Prob} in three simple steps and it is summarized in Algorithm \ref{my_algo}.

\begin{figure}[tb!]\centering
	\setlength\figureheight{5.6cm}
	\setlength\figurewidth{5.6cm}
    \subfloat[B\&B association.]{\label{fig:fig1a}
	\footnotesize
\definecolor{mycolor1}{rgb}{0.50000,1.00000,0.00000}%
\definecolor{mycolor2}{rgb}{0.00000,1.00000,1.00000}%
\definecolor{mycolor3}{rgb}{0.50000,0.00000,1.00000}%
\begin{tikzpicture}

\begin{axis}[%
width=\figurewidth,
height=\figureheight,
at={(0\figurewidth,0\figureheight)},
scale only axis,
xmin=0,
xmax=4000,
xlabel={X-coordinate (meters)},
xlabel near ticks,
xmajorgrids,
ymin=0,
ymax=4000,
ylabel={Y-coordinate (meters)},
ylabel near ticks,
ymajorgrids,
zmin=0,
zmax=400,
ztick={\empty},
zmajorgrids,
view={0}{90},
axis x line*=bottom,
axis y line*=left,
axis z line*=left,
]
\addplot[only marks,mark=o,mark options={},mark size=1.5000pt,color=black] plot table[row sep=crcr,]{%
555.275348242063	1269.81283677916\\
1333.83251528891	2834.54067266972\\
1971.20939847178	1402.11024365926\\
3948.02598839803	392.507773123991\\
3631.06125112028	2963.43065049088\\
2958.01881175789	1808.04090364484\\
180.858617636568	2316.32970881634\\
};

\addplot[only marks,mark=*,mark options={},mark size=2pt,color=red] plot table[row sep=crcr,]{%
3060.02039382597	3738.32804603065\\
2879.63114918649	3262.76963842476\\
1666.53827445954	2979.18522910982\\
3407.1180904332	3804.34989237501\\
3181.39438139178	3348.84589347927\\
};

\addplot3[only marks,mark=asterisk,mark options={},mark size=4.3301pt,color=red] plot table[row sep=crcr,]{%
2076.03940715119	3598.45497805639	300\\
};

\addplot[only marks,mark=diamond*,mark options={},mark size=3pt,color=mycolor1,fill=mycolor1] plot table[row sep=crcr,]{%
1444.03551474095	1497.90792055781\\
1418.41079363316	175.439112963146\\
373.08893023556	1732.78737770918\\
374.492120917132	759.212361303934\\
1509.78581459249	852.770513039731\\
};
\addplot3[only marks,mark=asterisk,mark options={},mark size=4.3301pt,color=mycolor1] plot table[row sep=crcr,]{%
1579.51462030496	598.443691232236	300\\
};
\addplot[only marks,mark=square*,mark options={},mark size=2pt,color=mycolor2,fill=mycolor2] plot table[row sep=crcr,]{%
3291.27389567016	1661.28088839505\\
2438.21924540486	2604.63794909011\\
2899.61179942553	1404.02020122884\\
2254.29690387144	1526.1136535509\\
3801.49935121189	1900.00713641047\\
};
\addplot3[only marks,mark=asterisk,mark options={},mark size=4.3301pt,color=mycolor2] plot table[row sep=crcr,]{%
3039.35849637011	1480.88749889574	300\\
};
\addplot[only marks,mark=triangle*,mark options={},mark size=2.5pt,color=mycolor3,fill=mycolor3] plot table[row sep=crcr,]{%
179.886089885807	3105.13745901857\\
715.857996032254	2965.20411152756\\
1343.31530776349	3245.61048461811\\
427.417258956023	2645.19748567078\\
317.127350381806	3834.03180602122\\
1035.51124830465	3062.63292345348\\
};
\addplot3[only marks,mark=asterisk,mark options={},mark size=4.3301pt,color=mycolor3] plot table[row sep=crcr,]{%
703.741819475246	3800.51506307402	300\\
};
\end{axis}
\end{tikzpicture}%
} \vfill \medskip
    \subfloat[Algorithm \ref{my_algo} association.]{\label{fig:fig1b}
	\footnotesize
\definecolor{mycolor1}{rgb}{0.50000,1.00000,0.00000}%
\definecolor{mycolor2}{rgb}{0.00000,1.00000,1.00000}%
\definecolor{mycolor3}{rgb}{0.50000,0.00000,1.00000}%
\begin{tikzpicture}

\begin{axis}[%
width=\figurewidth,
height=\figureheight,
at={(0\figurewidth,0\figureheight)},
scale only axis,
xmin=0,
xmax=4000,
xlabel={X-coordinate (meters)},
xlabel near ticks,
xmajorgrids,
ymin=0,
ymax=4000,
ylabel={Y-coordinate (meters)},
ylabel near ticks,
ymajorgrids,
zmin=0,
zmax=400,
ztick={\empty},
zmajorgrids,
view={0}{90},
axis x line*=bottom,
axis y line*=left,
axis z line*=left,
legend style={at={(-0.129307,-0.26949)},legend columns=-1,anchor=south west,legend cell align=left,align=left,draw=white!15!black}
]
\addplot[only marks,mark=o,mark options={},mark size=1.5000pt,color=black] plot table[row sep=crcr,]{%
555.275348242063	1269.81283677916\\
1333.83251528891	2834.54067266972\\
715.857996032254	2965.20411152756\\
1343.31530776349	3245.61048461811\\
427.417258956023	2645.19748567078\\
2879.63114918649	3262.76963842476\\
2438.21924540486	2604.63794909011\\
3407.1180904332	3804.34989237501\\
1035.51124830465	3062.63292345348\\
373.08893023556	1732.78737770918\\
3631.06125112028	2963.43065049088\\
};
\addlegendentry{Unassociated SBSs};

\addplot[only marks,mark=*,mark options={},mark size=2pt,color=red] plot table[row sep=crcr,]{%
3060.02039382597	3738.32804603065\\
1666.53827445954	2979.18522910982\\
3181.39438139178	3348.84589347927\\
};
\addlegendentry{Associated SBSs};

\addplot3[only marks,mark=asterisk,mark options={},mark size=4.3301pt,color=red] plot table[row sep=crcr,]{%
2076.03940715119	3598.45497805639	300\\
};
\addlegendentry{UAV-hubs};

\addplot[only marks,mark=diamond*,mark options={},mark size=3pt,color=mycolor1,fill=mycolor1] plot table[row sep=crcr,]{%
1444.03551474095	1497.90792055781\\
1971.20939847178	1402.11024365926\\
1418.41079363316	175.439112963146\\
374.492120917132	759.212361303934\\
1509.78581459249	852.770513039731\\
};
\addplot3[only marks,mark=asterisk,mark options={},mark size=4.3301pt,color=mycolor1] plot table[row sep=crcr,]{%
1579.51462030496	598.443691232236	300\\
};
\addplot[only marks,mark=square*,mark options={},mark size=2pt,color=mycolor2,fill=mycolor2] plot table[row sep=crcr,]{%
3291.27389567016	1661.28088839505\\
2899.61179942553	1404.02020122884\\
2254.29690387144	1526.1136535509\\
3801.49935121189	1900.00713641047\\
3948.02598839803	392.507773123991\\
2958.01881175789	1808.04090364484\\
};
\addplot3[only marks,mark=asterisk,mark options={},mark size=4.3301pt,color=mycolor2] plot table[row sep=crcr,]{%
3039.35849637011	1480.88749889574	300\\
};
\addplot[only marks,mark=triangle*,mark options={},mark size=2.5pt,color=mycolor3,fill=mycolor3] plot table[row sep=crcr,]{%
179.886089885807	3105.13745901857\\
317.127350381806	3834.03180602122\\
180.858617636568	2316.32970881634\\
};
\addplot3[only marks,mark=asterisk,mark options={},mark size=4.3301pt,color=mycolor3] plot table[row sep=crcr,]{%
703.741819475246	3800.51506307402	300\\
};
\end{axis}
\end{tikzpicture}%
}
	\caption{2D view of a random distribution and association of NFP-hubs and SBSs for $N_{BS}=28$, $N_D = 4$ and parameters defined in Table \ref{tab:SimPar}.}
	\label{fig:fig1}
\end{figure}
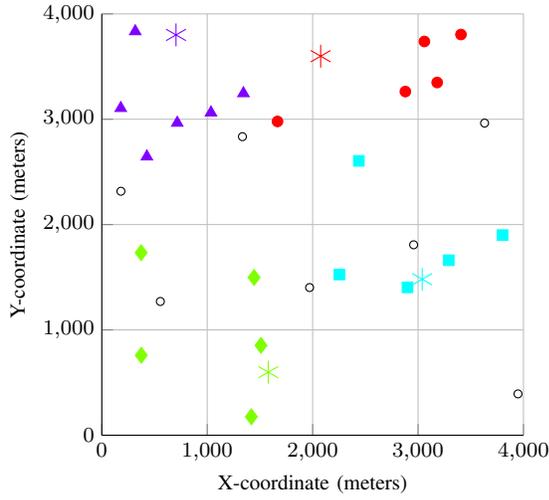
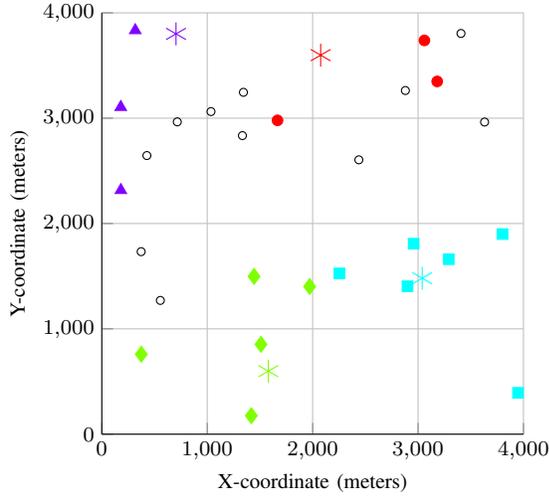

\section{Numerical Results}\label{sec:SimRes}

An urban square region with $\text{Area}=16$ km$^2$ is considered, where the SBSs are distributed using \emph{Matern} type-I hard-core process with an average density of $\lambda$ per $\text{m}^2$ having a minimum separation of $s_{BS}^{\text{min}}=300$ meters with each other. A snapshot of this distribution is considered and the best possible association is computed accordingly. Using the simulation parameters presented in Table \ref{tab:SimPar}, the number of NFPs $N_D$ from \eqref{NumDrones2} is computed and the distance that each NFP can cover $s_D^{\text{min}}$ from \eqref{PathLoss} is estimated. Then, the NFPs are distributed using \emph{Matern} type-I hard-core process with the same average density $\lambda$ but having a minimum separation of $s_D^{\text{min}}$ meters with each other. SBSs are assigned data rates randomly from the data rate vector $r_{\text{SBS}}$ shown in Table \ref{tab:SimPar} and the respective required bandwidth is computed as per simulation parameters.

In the considered case study, the average density $\lambda = 2 \times 10^{-6}$ per $\text{m}^2$ is used, that results in $N_{BS}=28$. \eqref{NumDrones1} and \eqref{NumDrones2} results in the number of NFP-hubs as $N_D = 4$. Fig. \ref{fig:fig1} shows the random distribution of both SBSs and NFP-hubs, where only 2D view of the region is shown as the NFP-hubs are at the same height $h_D = h_{D_{\text{max}}}=300$ meters. It can be noticed by the comparison of Fig. \ref{fig:fig1a} and Fig. \ref{fig:fig1b} that B\&B method results in more associated SBSs as compared to our proposed Algorithm \ref{my_algo}. However, serving more SBSs is not the primary objective of our optimization problem \eqref{eq:Opt_Prob} and thus, below the algorithms are further investigated on the basis of sum data rate. Moreover, it is clear from Fig. \ref{fig:fig1} that both algorithms are not able to associate the $N_{BS}$ SBSs with the NFP-hubs and the reason can be provided from Fig. \ref{fig:fig2}.

\begin{figure}[tb!]\centering
	\setlength\figureheight{5.1cm}
	\setlength\figurewidth{7.2cm}
	\footnotesize
\definecolor{mycolor1}{rgb}{1.00000,1.00000,0.00000}%
\definecolor{mycolor2}{rgb}{0.00000,0.44700,0.74100}%
\begin{tikzpicture}
\begin{axis}[%
width=\figurewidth,
height=\figureheight,
at={(0\figurewidth,0\figureheight)},
scale only axis,
area legend,
xmin=15,xmax=30,
xtick={17,21,22,28},
xlabel near ticks,
xlabel={No. of associated SBSs},
ymin=0, ymax=4,
ylabel={Sum rate (Gbps)},
ylabel near ticks,
legend style={at={(0.03,0.97)},anchor=north west,legend cell align=left,align=left,draw=white!15!black}
]
\addplot[ybar,bar width=0.06\figurewidth,bar shift=-0.018524\figurewidth,draw=black,fill=red,postaction={pattern=north east lines}] plot table[row sep=crcr] {%
17	1.98\\
};
\addlegendentry{Algorithm \ref{my_algo}};

\addplot[ybar,bar width=0.06\figurewidth,bar shift=-0.006175\figurewidth,draw=black,fill=green,postaction={pattern=north west lines}] plot table[row sep=crcr] {%
21	1.98\\
};
\addlegendentry{B\&B all};

\addplot[ybar,bar width=0.06\figurewidth,bar shift=0.006175\figurewidth,draw=black,fill=blue,postaction={pattern=horizontal lines}] plot table[row sep=crcr] {%
22	2.19\\
};
\addlegendentry{B\&B 3c to 3f};

\addplot[ybar,bar width=0.06\figurewidth,bar shift=0.018524\figurewidth,draw=black,fill=mycolor1,postaction={pattern=vertical lines}] plot table[row sep=crcr] {%
28	2.7\\
};
\addlegendentry{B\&B 3d to 3f};

\addplot [color=mycolor2,solid,forget plot]
  table[row sep=crcr]{%
15	2\\
30	2\\
};

\node[above, align=center, inner sep=0mm, text=black]
at (axis cs:16.7,2.1,0) {223.28};
\node[above, align=center, inner sep=0mm, text=black]
at (axis cs:20.7,2.1,0) {248.03};
\node[above, align=center, inner sep=0mm, text=black]
at (axis cs:22.3,2.3,0) {248.03};
\node[above, align=center, inner sep=0mm, text=black]
at (axis cs:28,2.8,0) {362.25};

\node at (axis cs:25.5,3.4,0) {\shortstack[l]{max. bandwidth\\of a NFP-hub (MHz)}};
\draw [->] (axis cs:25,3.18,0)--(axis cs:27,2.9,0);
\draw [->] (axis cs:25,3.18,0)--(axis cs:23,2.5,0);
\draw [->] (axis cs:25,3.18,0)--(axis cs:20.6,2.35,0);
\draw [->] (axis cs:25,3.18,0)--(axis cs:17.2,2.35,0);

\end{axis}
\end{tikzpicture}%
	\caption{Total sum data rate (Gbps) vs. the number of associated SBSs for B\&B method and Algorithm \ref{my_algo}.}
	\label{fig:fig2}
\end{figure}
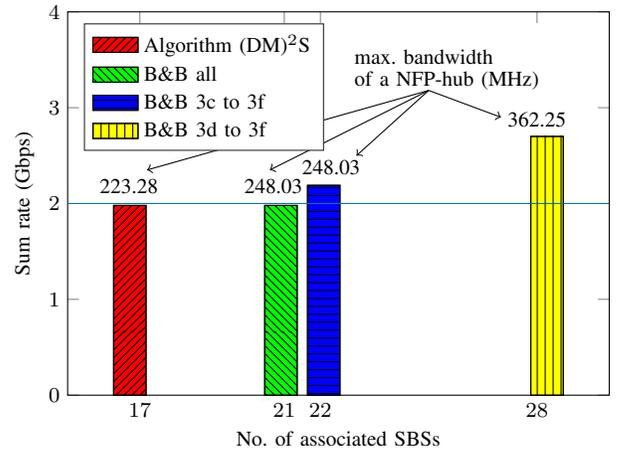

\begin{figure}[tb!]\centering
	\setlength\figureheight{5.1cm}
	\setlength\figurewidth{7.2cm}
	\footnotesize
\begin{tikzpicture}

\begin{axis}[%
ybar stacked,
width=\figurewidth,
height=\figureheight,
at={(0\figurewidth,0\figureheight)},
scale only axis,
bar width=0.063395\figurewidth,
area legend,
xmin=0,
xmax=15,
xtick={1,2,3,4,7,8,9,10},
xticklabels={{1},{2},{3},{4},{1},{2},{3},{4}},
xlabel={UAV-hub Number},
xlabel near ticks,
ymin=0,
ymax=7,
ytick={0,1,2,3,4,5,6,7},
ylabel={No. of Associated SBSs},
ylabel near ticks,
legend style={legend cell align=left,align=left,draw=white!15!black}
]
\addlegendimage{empty legend}
\addlegendentry{\hspace{-.6cm}SBSs with $r_{\text{SBS}}$}
\addplot+[ybar,postaction={pattern=vertical lines},fill=red,draw=black] plot coordinates {(1,1) (2,0) 
  (3,0) (4,1) (7,0) (8,0) (9,0) (10,0)};
\addlegendentry{30 Mbps};

\addplot+[ybar,postaction={pattern=north east lines},fill=blue,draw=black] plot coordinates {(1,1) (2,1) 
  (3,2) (4,2) (7,0) (8,0) (9,1) (10,0)};
\addlegendentry{60 Mbps};

\addplot+[ybar,postaction={pattern=north west lines},fill=green,draw=black] plot coordinates {(1,0) (2,3) 
  (3,0) (4,1) (7,0) (8,3) (9,2) (10,0)};
\addlegendentry{90 Mbps};

\addplot+[ybar,postaction={pattern=grid},fill=yellow,draw=black] plot coordinates {(1,1) (2,0) 
  (3,2) (4,2) (7,1) (8,0) (9,2) (10,3)};
\addlegendentry{120 Mbps};

\addplot+[ybar,postaction={pattern=dots},draw=black] plot coordinates {(1,2) (2,1) 
  (3,1) (4,0) (7,2) (8,2) (9,1) (10,0)};
\addlegendentry{150 Mbps};

\node at (25,650) {B\&B method};
\node at (75,620) {\shortstack[l]{Algorithm \\ \ref{my_algo}}};

\end{axis}
\end{tikzpicture}%
	\caption{Number of associated SBSs vs. every NFP-hub 1-4 for both B\&B method and Algorithm \ref{my_algo}.}
	\label{fig:fig3}
\end{figure}
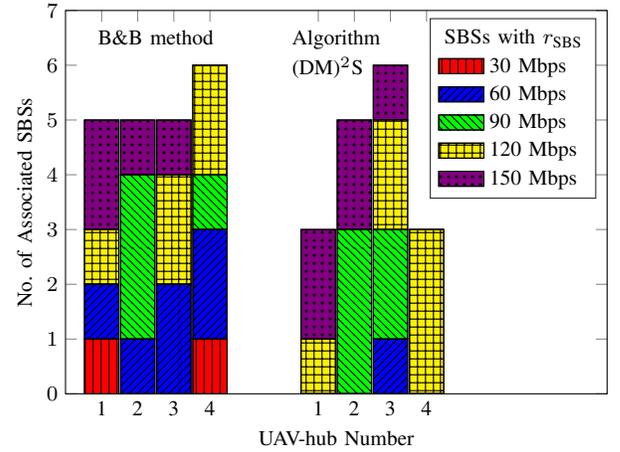

Fig. \ref{fig:fig2} plots the sum data rate vs. the number of associated SBSs considering the same distribution and parameters as in Fig. \ref{fig:fig1}. It can be seen that if only constraints \eqref{cons3} to \eqref{cons5} are considered, then B\&B algorithm results in association of all available SBSs with the NFP-hubs. However, in this case the association exceeds the backhaul data rate $R=2$ Gbps and NFP bandwidth $B=250$ MHz limits shown in Table \ref{tab:SimPar}. Then, if other constraints are considered too, the algorithm results in a different number of associated SBSs. This shows that all these constraints, that are not considered in related literature, affects the association. By comparing the results of Algorithm \ref{my_algo} with the B\&B method considering all the constraints, it can be noticed that both provides same sum data rate and thus have the same performance. Further, both these solutions satisfy the maximum bandwidth constraint \eqref{cons2}.

To investigate reason for less associated SBSs by Algorithm \ref{my_algo} as compared to B\&B method, Fig. \ref{fig:fig3} shows the number of associated SBSs vs. every NFP-hub 1-4 for both methods considering all constraints. It can be noticed that both these solutions satisfy the constraint \eqref{cons4}, i.e., $N_l=7$. Fig. \ref{fig:fig3} shows that our algorithm gives priority to SBSs demanding high data rate, which agrees with the design of Algorithm \ref{my_algo}, and thus, it results in less number of associated SBSs. Note that, with appropriate modifications in Step 2 and 3 of the Algorithm \ref{my_algo}, more SBSs can be associated. However, the objective here is just to maximize the sum rate, which Algorithm \ref{my_algo} satisfies as shown in Fig. \ref{fig:fig2}.

\section{Computational Complexity}\label{sec:Comp_Complexity}
Both the B\&B method and the algorithm \ref{my_algo} are self-learning algorithms and therefore, it is difficult to provide average performance tight bound comparison for them. Thus, here, the worst case computational complexity of both algorithms is discussed and then the average run time speed is analyzed. It is well known in literature that the worst case performance bound of the B\&B method is same as of Brute-force \cite{BandB_Algo} and \cite{zhang1996branch}. The computational complexity of B\&B method and Algorithm \ref{my_algo} is compared in terms of the number of flops in Table \ref{CompComplex} for the worst case scenario. It can be noticed from Table \ref{CompComplex} that the Algorithm \ref{my_algo} is cheaper than Brute-force and thus the B\&B method in the worst case and provides the same performance as can be observed from the simulation results.

\begin{table}[tb!]
\renewcommand{\arraystretch}{1.5}
\centering
\caption{Computational complexity of B\&B and \ref{my_algo} Algorithms.}
\label{CompComplex}
\begin{tabular}{|c|c|}
\hline
\textbf{Algorithm}  &  \textbf{Complexity Order}     \\ \hline
Brute-force         &  $N_D^3 N_{BS}^{N_D + 2}$      \\ \hline
\ref{my_algo}       &  $N_{BS} N_D + \mathcal{O}\left(N_{BS} \left( N_l + N_D \right)\right)$   \\ \hline
\end{tabular}
\end{table}

\begin{table}[tb!]
\renewcommand{\arraystretch}{1.2}
\centering
\caption{Run Time comparison of B\&B method and Algorithm \ref{my_algo}.}
\label{tab:TimComp}
\begin{tabular}{|c|c|c|}
\hline
\textbf{Method}          & \textbf{Sum rate (Gbps)} & \textbf{Elapsed Time (seconds)}   \\ \hline
B\&B                     & 1.98                     & 2.3373                            \\ \hline
Algorithm \ref{my_algo}  & 1.98                     & 0.0315                            \\ \hline
\end{tabular}
\end{table}

Table \ref{tab:TimComp} compares our algorithm \ref{my_algo} with the B\&B method in terms of the overall sum rate of the network and elapsed time of the algorithm to solve the optimization problem \eqref{eq:Opt_Prob} for the case study presented in Section \ref{sec:SimRes}. It can be noticed that both has the same performance in terms of the sum rate, however, Algorithm \ref{my_algo} achieves this result in much smaller run-time duration. Therefore, it can be said that our proposed method is computationally less expensive not only in terms of worst case analysis but also in average runtime analysis and thus, it is practically applicable. Note that, both algorithms are implemented on MATLAB R2014b on a Windows 8 platform running over a machine with core i5 processor clocked at 2.5 GHz with 4 GB RAM.

\balance
\section{Conclusions}\label{sec:Conc}
This paper considers the use of NFP-hubs to provide connectivity to SBSs with the core network. An optimization problem is formulated for their association considering backhaul data rate limitation and a number of NFP related limitations such as the maximum number of supported links and bandwidth. Our proposed distributed algorithm named as Distributed Maximal Demand Minimum Servers ((DM)$^2$S), performs a greedy search on the basis of maximum SINR links and gives priority to SBSs demanding high data rate in order to maximize the overall sum rate of the network. Numerical evaluation of a case study has shown a favourable performance of our proposed algorithm as compared to exhaustive B\&B method and because of its lower computational complexity and distributive nature, it can be practically implemented. Here, for brevity, only a single case study is considered to compare and contrast our proposed algorithm with the B\&B method. In the future, we investigate further the performance of proposed algorithm and look for the needed enhancements in various cases such as the network centric case \cite{ElhamBackhaul}, where focus is to serve maximum possible SBSs instead of giving priority to the ones demanding high data rate that we considered in this work.

\bibliographystyle{IEEEtran}
\bibliography{UAV_hub_BS}

\begin{thebibliography}{10}
\providecommand{\url}[1]{#1}
\csname url@samestyle\endcsname
\providecommand{\newblock}{\relax}
\providecommand{\bibinfo}[2]{#2}
\providecommand{\BIBentrySTDinterwordspacing}{\spaceskip=0pt\relax}
\providecommand{\BIBentryALTinterwordstretchfactor}{4}
\providecommand{\BIBentryALTinterwordspacing}{\spaceskip=\fontdimen2\font plus
\BIBentryALTinterwordstretchfactor\fontdimen3\font minus
  \fontdimen4\font\relax}
\providecommand{\BIBforeignlanguage}[2]{{%
\expandafter\ifx\csname l@#1\endcsname\relax
\typeout{** WARNING: IEEEtran.bst: No hyphenation pattern has been}%
\typeout{** loaded for the language `#1'. Using the pattern for}%
\typeout{** the default language instead.}%
\else
\language=\csname l@#1\endcsname
\fi
#2}}
\providecommand{\BIBdecl}{\relax}
\BIBdecl

\bibitem{Paradigm_Andrew}
J.~G. Andrews, ``Seven ways that {HetNets} are a cellular paradigm shift,''
  \emph{IEEE Commun. Magazine}, vol.~51, no.~3, pp. 136--144, Mar. 2013.

\bibitem{5G_Andrew}
J.~G. Andrews, S.~Buzzi, W.~Choi, S.~V. Hanly, A.~Lozano, A.~C.~K. Soong, and
  J.~C. Zhang, ``What will {5G} be?'' \emph{IEEE J. on Sel. Areas in Commun.},
  vol.~32, no.~6, pp. 1065--1082, Jun. 2014.

\bibitem{robson2012backhaul}
J.~Robson, ``Small cell backhaul requirements,'' \emph{NGMN White Paper}, pp.
  1--40, 2012.

\bibitem{Fiber_Backhaul}
C.~Ranaweera, M.~G.~C. Resende, K.~Reichmann, P.~Iannone, P.~Henry, B.~J. Kim,
  P.~Magill, K.~N. Oikonomou, R.~K. Sinha, and S.~Woodward, ``Design and
  optimization of fiber optic small-cell backhaul based on an existing
  fiber-to-the-node residential access network,'' \emph{IEEE Commun. Magazine},
  vol.~51, no.~9, pp. 62--69, Sep. 2013.

\bibitem{FSO_Ter_backhaul}
Y.~Li, M.~Pióro, and V.~Angelakisi, ``Design of cellular backhaul topology
  using the {FSO} technology,'' in \emph{2nd IEEE IWOW Workshop}, pp. 6--10,
  Oct. 2013.

\bibitem{ShakirFSOMAG}
M.~Alzenad, M.~Z. Shakir, H.~Yanikomeroglu, and M.-S. Alouini, ``{FSO}-based
  vertical backhaul/fronthaul framework for {5G+} wireless networks,''
  \emph{arXiv preprint arXiv:1607.01472}, 2016.

\bibitem{ATGmodel}
A.~Al-Hourani, S.~Kandeepan, and A.~Jamalipour, ``Modeling air-to-ground path
  loss for low altitude platforms in urban environments,'' in \emph{IEEE
  GLOBECOM}, pp. 2898--2904, Dec. 2014.

\bibitem{ATG_optDrone1}
A.~Al-Hourani, S.~Kandeepan, and S.~Lardner, ``Optimal {LAP} altitude for
  maximum coverage,'' \emph{IEEE Wireless Commun. Lett.}, vol.~3, no.~6, pp.
  569--572, Dec. 2014.

\bibitem{TwoDrones}
M.~Mozaffari, W.~Saad, M.~Bennis, and M.~Debbah, ``Drone small cells in the
  clouds: Design, deployment and performance analysis,'' in \emph{IEEE
  GLOBECOM}, pp. 1--6, Dec 2015.

\bibitem{IremOneDrone}
R.~I. Bor-Yaliniz, A.~El-Keyi, and H.~Yanikomeroglu, ``Efficient {3-D}
  placement of an aerial base station in next generation cellular networks,''
  in \emph{IEEE ICC}, pp. 1--5, May. 2016.

\bibitem{ElhamBackhaul}
\BIBentryALTinterwordspacing
E.~Kalantari, M.~Z. Shakir, H.~Yanikomeroglu, and A.~Yonga{\c{c}}oglu,
  ``Backhaul-aware robust {3D} drone placement in {5G+} wireless networks,''
  \emph{arXiv preprint arXiv:1702.08395}, Mar. 2017. [Online]. Available:
  \url{http://arxiv.org/abs/1702.08395}
\BIBentrySTDinterwordspacing

\bibitem{ElhamMultiPSO}
E.~Kalantari, H.~Yanikomeroglu, and A.~Yongacoglu, ``On the number and {3D}
  placement of drone base stations in wireless cellular networks,'' in
  \emph{IEEE VTC}, pp. 1--6, Sep. 2016.

\bibitem{Mozaffari2016}
M.~Mozaffari, W.~Saad, M.~Bennis, and M.~Debbah, ``Efficient deployment of
  multiple unmanned aerial vehicles for optimal wireless coverage,'' \emph{IEEE
  Commun. Lett.}, vol.~20, no.~8, pp. 1647--1650, Aug. 2016.

\bibitem{Sharma2016}
V.~Sharma, K.~Srinivasan, H.-C. Chao, K.-L. Hua, and W.-H. Cheng, ``Intelligent
  deployment of {UAVs} in {5G} heterogeneous communication environment for
  improved coverage,'' \emph{J. of Network and Computer Applications}, pp.~--,
  2016.

\bibitem{matern2013}
B.~Mat{\'e}rn, \emph{Spatial variation}, ser. Springer Lecture Notes in
  Statistics.\hskip 1em plus 0.5em minus 0.4em\relax Springer, 1986, vol.~36.

\bibitem{NPhard1}
E.~Karamad, R.~S. Adve, Y.~Lostanlen, F.~Letourneux, and S.~Guivarch,
  ``Optimizing placements of backhaul hubs and orientations of antennas in
  small cell networks,'' in \emph{IEEE ICCW}, pp. 68--73, Jun. 2015.

\bibitem{NPhard2}
Z.~Mlika, E.~Driouch, W.~Ajib, and H.~Elbiaze, ``A completely distributed
  algorithm for user association in {HetSNets},'' in \emph{IEEE ICC}, pp.
  2172--2177, Jun. 2015.

\bibitem{BandB_Algo}
A.~Schrijver, \emph{Theory of linear and integer programming}.\hskip 1em plus
  0.5em minus 0.4em\relax New York, NY, USA: John Wiley \& Sons, 1986.

\bibitem{zhang1996branch}
W.~Zhang, ``Branch and {Bound} search algorithms and their computational
  complexity.'' DTIC Document, Tech. Rep., 1996.

\end{thebibliography}

\end{document}